# Design Diversity for Improving Efficiency and Reducing Risk in Oil and Gas Well Stimulation under Uncertain Reservoir Conditions

C. Cheng[*]

*Department of Chemical and Petroleum Engineering, University of Pittsburgh, Pittsburgh, PA, USA*

## Abstract

Hydraulic fracturing stimulates fracture swarm in reservoir formation though pressurized injection fluid. However restricted by the availability of formation data, the variability embraced by reservoir keeps uncertain, driving unstable gas recovery along with low resource efficiency, being responsible for resource scarcity, contaminated water and injection-induced seismicity. Resource efficiency is qualified though new determined energy efficiency, a scale of recovery and associated environmental footprint. To maximize energy efficiency while minimize its' variation, we issue picked designs at reservoir conditions dependent optimal probabilities, assembling high efficiency portfolios and low risk portfolios for portfolio combination, which balance the variation and efficiency at optimal by adjusting the proportion of each portfolio. Relative to regular design for one well, the optimal portfolio combination applied in multiple wells receive remarkable variation reduction meanwhile substantial energy efficiency increase, in response to the call of more recovery per unit investment and less environment cost per unit nature gas extracted.

## 1. Introduction

Hydraulic fracturing (HF), a well stimulation technique in which rock is fractured by pressurized liquid, plays increasingly important role in energy production. Enabled by hydraulic fracturing, shale gas as a clean energy, has supplied about 75% of total U.S. dry natural gas production in 2019 (1). However associated environmental effect is non-negligible (2-4). The environmental footprints are scaled by the efficiency of resource usage (5-7), which is controlled by the variability and uncertainty of HF. The variability could be identified as the naturally possessed one due to the sedimentation, tectonic movement (8-9) and the human induced one due to the fracturing (10-13). Based on existing technologies, the variability partially transfers to be uncertainty because it is still challengeable to get formation variation accurately and economically impossible to measure these parameters well by well. An uncertain output is not out of expect considering that the variations remain uncertain not only between neighbor clusters and also multiple wells. A widely used method is the extreme limited entry (EXL), dominating the global pressure to overcome the uncertain stress distribution though high injection pressure (14-16). To ensure the pressure could be compatible with the uncertainty of the in-situ stress, up to 80MPa injection pressure is applied. Massive resource such as the injection fluid and energy are left in place, leading to environmental, monetary and societal cost, including pollution (17-19), resource waste (20,21) and suspected earthquake (22-24).

Rather than pursuing single optimal design, portfolio combination is invoked and identified as a promising way to manage the uncertainty caused low efficiency. In detail, distinctive designs are applied at certain probability of proceeding wells to stabilize the enhanced resource efficiency of whole operating area, instead of repeating one design. Regarding each design is made up of five variables: perforation pressure, viscosity, stage length, injection rate and spacing, in turn letting the optional designs countless. To focus on the most potential designs, energy efficiency is introduced as a qualification of resource efficiency, which is calculated though the portion of energy account for fracturing in total input energy, involving the injection fluid volume, pressure applied and fractured area. The deviation of energy efficiency between clusters is defined as energy variance. Lower energy variance promotes more evenly distributed proppants, meaning more evenly propped fractured area and more productive. Because proppants hold fractures open to enable nature gas flow more freely through the fractures when hydraulic pressure is removed from the well. The frontier of energy efficiency and energy variance is figured out and then chosen from that to get distinctive design with design parameters ready for uncertain formation variation, which is qualified though the randomly valued young's modules, toughness, leak-off coefficient and in-situ stress.

To account for the uncertain output driven by varied parameters, risk is determined to qualify the deviation of both energy efficiency and energy variance. Energy efficiency is set as the return. The design portfolio is constructed by mixing the picked-up designs with associated return and risk at random weight. Inspired by 'Modern portfolio theory (MPT)' from economy (25), the return and risk of the design portfolio is calculated out. A clear frontier of efficiency and risk is achieved. It is observed that the portfolio could approach similar risk with single extreme limited entry but times more energy efficiency. Motivated by the target of risk reduction, we find that if the mixture process is also applied on the EXL designs, another 50% lowered risk is offered. Multiple lines are projected from these EXL portfolios to high efficiency but risky portfolios. The points along the lowest gradient line are the optimal portfolio combination, a merged of high efficiency and low risk portfolios, owning the best tradeoff between energy efficiency and risk. Compare with repeated base design, the optimal portfolio combination can stimulate up to 250% more energy efficiency and reduce up to 80% risk. The results show that owning different kinds of designs at refined probability, could leverage the benefit and disadvantage of each design, provide a new thought to cover both environmental and economical concern.

## 2. Background

Encountered with formation variability during HF process is inevitable. To overcome the variability from stress, extreme limited entry (EXL) is used to suppress the stress variation by dominating the pressure distribution. If in-situ stress is 30Mpa, 10% variability will induce a 3Mpa stress variation, meaning that at least additional 3Mpa perforation pressure required. While to ensure the superiority over the in-situ stress , a perforation pressure from 3Mpa to 80Mpa is widely used in real operation. No doubt that there is huge energy wasted and great resource left behind like water, thickening agents and proppants during the extreme limited entry. As seen variability is the main issue leading to high energy and resource consumption. In our previous work, a designed spacing between fractures with an optimized fluid viscosity has been figured out to be the potential option to overcome the variability, in place of the EXL with lower input pressure and higher resource efficiency (26). However these specifically optimized methods work effectively only for stress variability. In practice, the other formation variables such as the elasticity, toughness and leak-off coefficient are also varied along the well and radial direction, which is so challengeable to be detected with enough accuracy considering that some could be measurable only though the core and technically unmeasurable in normal surface. Besides that, the variability also exists between wells. There are hundreds well drilling at same reservoir, the design for single well will perform differently from well to well. All of these induce one result: uncertain return with a high deviation, harmful for resource efficiency.

    Promised by the large scale randomly distributed formation uncertainty, the uncertainty in single well could be regarded as no difference with the uncertainty among multiple wells. Compare to passively suffering the uncertain effect by repeating single design, a design portfolio constituted by distinctive designs with corresponding probabilities provides a chance to leverage uncertainty

by absorbing benefit and relieving risk among designs. The designs could be picked up from thousands simulated designs enabled by C5Frac model, which is validated with several high-fidelity models but requires order's short computation time (26). Instead of optimizing particular design factor for one well with great efforts, optimizing the probability of each design to maximize the resource efficiency with minimized variation of multiple wells comes into thought. The performance evaluation is inspired by Modern portfolio theory (MPT) (25), where along the left boundary of the portfolios is the minimal risk for a given return, called efficient frontier Figure 1. A fixed rate of interest without variation is promised by deposit account, called risk-free rate shown as a cross point on y-axis. The tangent line projected from the risk-free rate to the efficient frontier is capital market link (CML) (27). The tangency point represents a portfolio with 100% of holdings. The points on the CML line bellow tangency point is a portfolio containing both the risk-free asset and the tangency portfolio. Such that the expected return is maximized for a given level of risk. Based on the assumption that investors are risk averse, meaning that given two portfolios that offer the same expected return, investors will prefer the portfolio on the CML line.

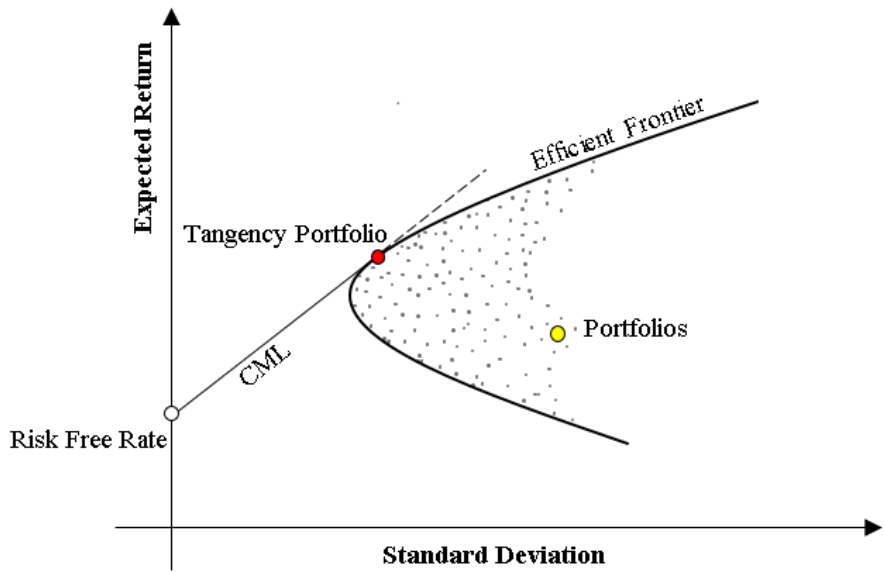

**Fig. 1. Illustration of MPT and CML.** The black points are a cloud of portfolios with one randomly picked point highlighted as yellow marked and named. The parabolic curve is the frontier of the portfolios called efficient frontier. With the risk-free asset, the straight line referred to CML, is tangent to the efficient frontier. The contact red point represents the tangency portfolio shown as a red spot.

## 3. Methods

Assembling the portfolio combination by mix all HF designs is obviously unpractical. Thus, scope reduction become necessary in first step. To seek the potential designs for mixture, energy efficiency is determined to qualify how much percent energy is used into fracturing. Higher energy efficiency means higher resource usage efficiency. Energy variance is determined to qualify the uniformity of fracturing energy, implicitly in duty of proppant distribution which leave space for oil and gas outflow. Lower deviation indicates more energy production due to more uniformly propped cracks. The detail is shown in Section S2. To detect the effect of spacing between interferential clusters, a fixed uniform spacing ratio ($h_1/Z$) with other four random design parameters: perforation pressure, injection rate, viscosity and stage length is input into our validated and $10^3$ times faster model C5Frac and this random simulations run thousands time. As a contrast, all five design parameters are randomly picked up in another thousands simulation runs. Table S1 provides a range of parameters in which that design might give rational energy efficiency and variance. The random sampling of each parameter utilizes Latin hypercube to reach all parametric space as efficiently as possible. Multiple dimensions' sampling contained in each design make the design picks are completely uncorrelated.

The energy variance versus energy efficiency is plotted in Figure 2(a). We find that the energy variance always grows with higher energy efficiency no matter the spacing. We cannot get both of them at optimal. So we pick designs along the frontier from $2.5*10^{-4}$ to respective max efficiency at $0.1*10^{-4}$ interval, obtaining minimum energy variance at given energy efficiency Figure 2(a). Design picks strictly follow same rule for uniform and non-uniform spacing. There are six designs picked along the frontier of uniform spacing, while there are twelve optimal designs available in non-uniform spacing. This is due to that non-uniform spacing could adjust the nonuniform stress distribution, asking less perforation pressure with 30% more energy efficiency Figures S2(c) and (d). The advantage over uniform spacing is also shown in 80% less energy variance. In reality, the design performance will be affected by the formation variation. The optimization for one well is not optimal for another well.

To account for the uncertainty possessed variation, uncertain level is defined as the dimensionless variation range of the formation parameters, which could be roughly estimated though just two wells and more accurately by multiple wells. Then the unknown formation variation will be qualified as a random value limited by the upper and bottom value provided by uncertain level. We apply same uncertainty level on in-situ tress, elasticity and toughness, while larger on leak-off coefficients with log scale because that permeability and porosity varied in several orders. For 5% uncertainty, the variation range is from $10^{-2.5\%}$ to $10^{2.5\%}$ for leak-off coefficient but 97.5% to 102.5% for other parameters. The values are independently randomly picked and uniformly distributed over that range for each case though Latin hypercube. Acknowledge that other distributions such as Cauchy distribution may somewhat effect the result, but the principle of this paper will not change.

With the random variation (Table S2), the previously picked designs will be individually put into C5Frac to compute the deviation of the returned energy efficiencies and energy deviation. , The summation of these two deviations is set as the risk and the mean value of energy efficiency is set as the return for each design. Assembling designs at random probability leads to thousands design portfolios, of which the risk and return is calculated though MPT, a mathematical framework (25) to mix individual asset with associated specific risk $\sigma_i$, return $r_i$ and the weighting $\rho_i$ of component asset $i$ (that is, the proportion of asset "$i$" in the portfolio). The $r$ and $\rho$ is given by

$$E(r) = \begin{bmatrix} r_1 \\ \vdots \\ r_N \end{bmatrix}, \quad \rho = \begin{bmatrix} \rho_1 \\ \vdots \\ \rho_N \end{bmatrix}, \quad \sum_{i=1}^{N} \rho_i = 1 \tag{1}$$

The expected return of portfolio is given by:

$$E(r_p) = \rho' E(r) = [\rho_1 \cdots \rho_N] \begin{bmatrix} r_1 \\ \vdots \\ r_N \end{bmatrix} = \sum_{i=1}^{N} \rho_i r_i \tag{2}$$

To cover the strength of the correlation between two or more sets of random variates. The covariance for two random assets $i$ and $j$ in portfolio, each with risk $\sigma$, is defined by

$$cov(i,j) = \begin{bmatrix} \sigma_{11} & \cdots & \sigma_{1N} \\ \vdots & \ddots & \vdots \\ \sigma_{N1} & \cdots & \sigma_{NN} \end{bmatrix}, \sigma_{ij} = \sigma_i \sigma_j \tag{3}$$

Portfolio return variance is given by:

$$\sigma_p^2 = \rho' cov(i,j)\rho = [\rho_1 \cdots \rho_N] \begin{bmatrix} \sigma_{11} & \cdots & \sigma_{1N} \\ \vdots & \ddots & \vdots \\ \sigma_{N1} & \cdots & \sigma_{NN} \end{bmatrix} \begin{bmatrix} \rho_1 \\ \vdots \\ \rho_N \end{bmatrix} = \sum_{i=1}^{N}\sum_{j=1}^{N} \rho_i \rho_j \sigma_{ij} \tag{4}$$

Because the designs we picked are fully independent, so the elements $\sigma_{ij}$ is 0 except for the diagonal elements $\sigma_{ii}$. Mixing designs in weights $\rho_i$ is the so-called portfolio, of which the return $E(r_p)$ and risk $\sigma_p$ could be calculated out based on Equations 2 and 4. Here we do mixture of six designs randomly chosen from picked designs with random probabilities from 0% to 100% though Monte Carlo method. The random mixing runs $1.25*10^4$ times to ensure the diversity of portfolios enough for cases coverage. As a compare, we choose another six designs from the frontier simulated in same uncertainty level, but for repetition in all wells. The design picks are from $2*10^{-4}$ to $2.5*10^{-4}$ at $0.1*10^{-4}$ interval shown as cyan points in Figure 2(a).

## 4. Results

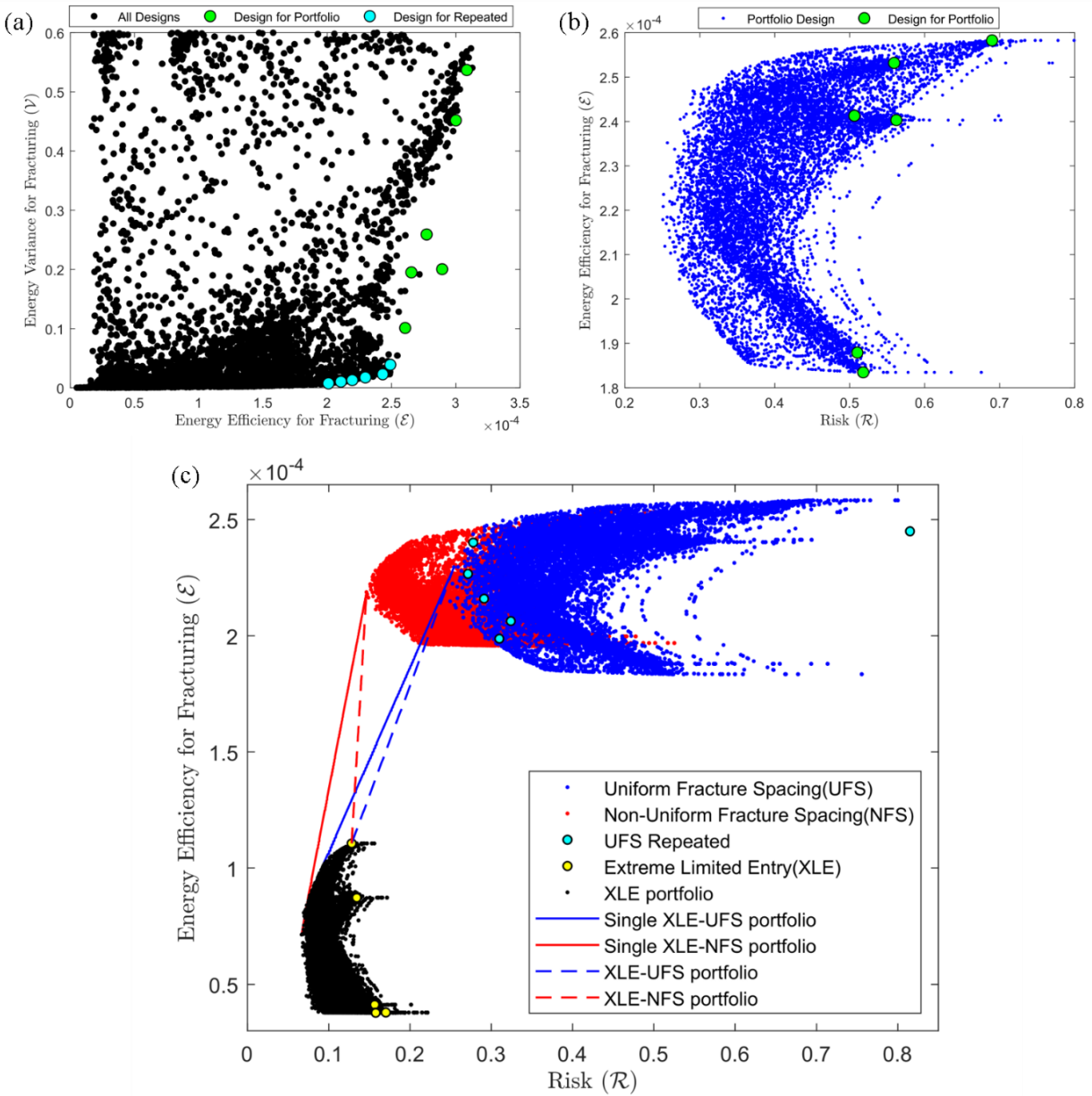

**Fig. 2. Compare portfolio of designs with repeated single design.** (**a**) The evolution of energy variance with energy efficiency. The green points are chosen for mixture. The cyan points are picked for repeat. (**b**) Assigned formation variation, the performance of each picked design before mixture is indicated by the green points. The blue points are the design portfolios. (**c**) The repeated

designs are shown with the portfolios. The dashed lines are drawn from EXL and the solid lines are drawn from portfolio of EXL to UFS and NFS portfolio individually.

We observe that if energy efficiency is higher than a certain value $2.5*10^{-4}$, energy variance will be substantially higher with a clear surged frontier Figure 2(a). Compare to individual design at same expected energy efficiency, a significant risk reduction is obtained in portfolio. Even for most risky highest efficiency design, 40% risk could be deducted at 2% cost of energy efficiency Figure 2(b). The designs for repetition give rise 110% more risk than the portfolio of non-uniform spacing with no observed advantage Figure 2(c). Extra risk reducing promoted by non-uniform spacing in 5% uncertain level, is suppressed in 20% uncertainty (see Figure S6), where stress variation is dominant by in-situ stress instead of interaction stress. To reach lower risk, six designs with lowest risk are chosen from extreme limited entry and running random mixture to assemble EXL portfolios, shown as the black points in Figure 2(c). Compare to the risk of single EXL, the portfolios could contribute additional 15% risk reduction. Tangent lines are drawn from EXL portfolios to the frontier of high efficiency portfolios. The line with lowest gradient balances the risk and return at lowest risk cost to progress energy efficiency. The points presented on this line are optimal portfolio combination, the optimal mixtures of tangent high efficiency portfolio and corresponding low risk portfolio for requested efficiency or limited risk. The change in efficiency is linearly related to the change in risk as the portfolio proportion vary. Any other point below this line will induce lower efficiency at same risk and higher risk at same efficiency. For same energy efficiency $2*10^{-4}$, optimal portfolio combination could halve the risk of NFS frontier portfolio.

Having established the method to achieve portfolio combinations, we turn to study the combinations at different uncertain level and the design contained in mixed portfolio. Three uncertainty levels are chosen: 5%, 10% and rarely seen 20% to cover most cases, being helpful to understand the role of uncertainty play in general. The corresponding variation is detailed in Table S2. A base case is divided by the optimal combinations to give the dimensionless risk and energy efficiency, enabling the compare of different uncertain level in same dimension. The base case is a typical design come with $0.2m^3$ injection rate, 50m stage length, 0.003Pa.s fluid viscosity and uniform spacing. The perforation factor is set as $1.06*10^{10}Pa.s^2/m^6$ to ensure the pressure of entry loss is around $1.38*10^7Pa$ (equal 2000Psi). The connected two points marked at either end of the line is the portfolios ready for mixture to balance the risk and return at best tradeoff. High efficiency portfolio located at right top and left bottom represents low risk portfolio. The probability of each design observed in the portfolio forms probability clouds displayed by the hexagram for better visualization. The designs are named with probability shown at the bottom of Figure 3 and corresponding design parameters listed in Table 1.

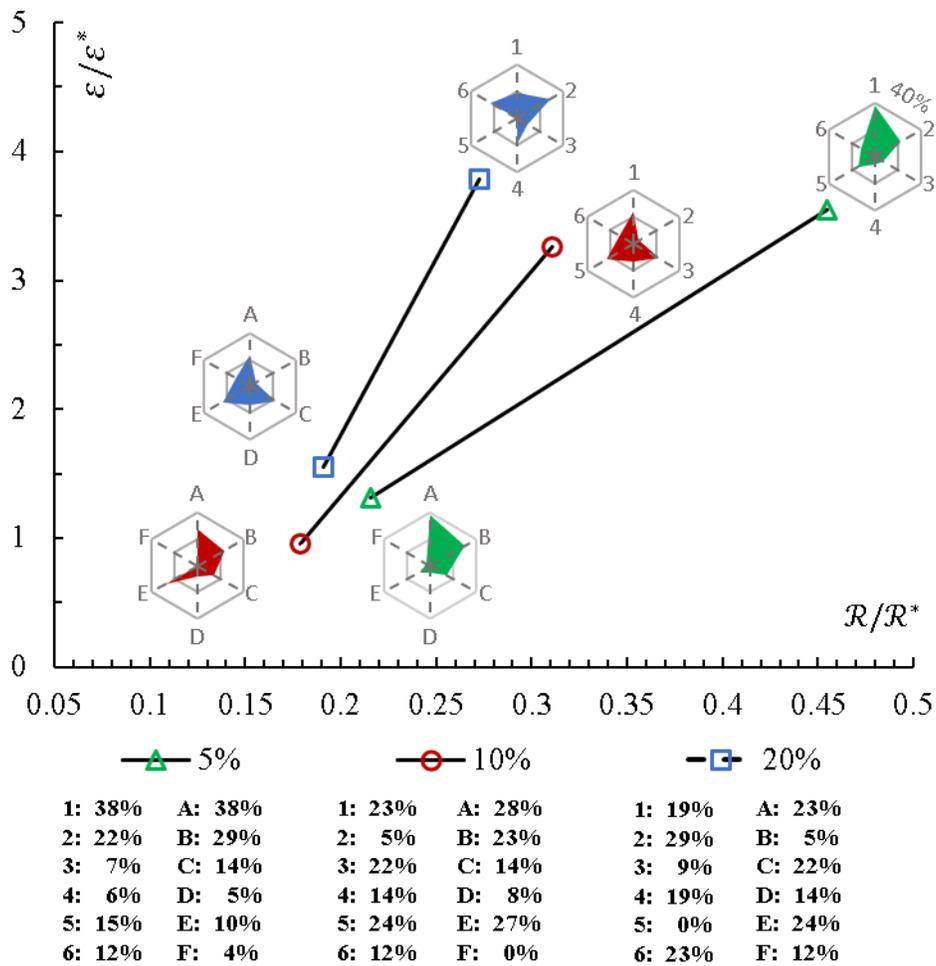

**Fig. 3. Optimal portfolio combinations at different uncertainty levels.** The vertical axis is the dimensionless energy efficiency and horizontal axis is dimensionless risk. The line with green triangle represents the combination of high efficiency portfolio and low risk portfolio at 5% uncertainty level. The line marked with red circle demonstrated the combination 10% uncertain level and 20% uncertain level comes with blue square. The probability of each design named after A to F for less risky portfolio and one to six for high efficiency portfolio. The probability of each design is list at the bottom.

Optimal portfolio combinations, arisen from the mixture between portfolios, could reduce 55% to 79% risk with three times energy efficiency at 5% uncertainty, while enlarged to 73%-82% at 20% uncertainty. Exact benefit of the combination is dependent on either anticipated efficiency or risk tolerance. According to the gradient, the combination of portfolios is more necessary in 5% uncertain level than in 10% or 20%, owing to more risk reduced at same efficiency cost. The corresponding design parameters contained in high efficiency portfolio (Figure 3) is illustrated in Tables 1.

**Table 1. Portfolio Designs at 5%, 10% and 20% Uncertainty**

| Uncertain Level | 5% | | | | | |
|---|---|---|---|---|---|---|
| Design | 1 | 2 | 3 | 4 | 5 | 6 |
| $h_1/Z$ (spacing ratio) | 0.3566 | 0.3497 | 0.3494 | 0.3568 | 0.3503 | 0.3606 |
| $\mu$ (viscosity Pa.s) | 0.544375 | 0.205375 | 0.171967 | 0.213425 | 0.200125 | 0.569617 |
| $Z$ (stage length m) | 20.55 | 30.44 | 27.95 | 31.3 | 28.57 | 20 |
| $Q_o$ (injection rate $m^3/s$) | 0.10329 | 0.12533 | 0.17493 | 0.1225 | 0.15599 | 0.10993 |
| $p_{perf}$ (perforation factor $Pa.s^2/m^6$) | $1.3*10^9$ | $3.1*10^6$ | $1.5*10^6$ | $1.2*10^7$ | $6.6*10^5$ | $8.4*10^8$ |

| Uncertain Level | 10% | | | | | |
|---|---|---|---|---|---|---|
| Design | 1 | 2 | 3 | 4 | 5 | 6 |
| $h_1/Z$ (spacing ratio) | 0.3487 | 0.3432 | 0.3494 | 0.3568 | 0.3503 | 0.3497 |
| $\mu$ (viscosity Pa.s) | 0.1756 | 0.429667 | 0.171967 | 0.213425 | 0.200125 | 0.205375 |
| $Z$ (stage length m) | 24.79 | 30.45 | 27.95 | 31.3 | 28.57 | 30.44 |
| $Q_o$ (injection rate $m^3/s$) | 0.15349 | 0.10977 | 0.17493 | 0.1225 | 0.15599 | 0.12533 |
| $p_{perf}$ (perforation factor $Pa.s^2/m^6$) | $1.4*10^6$ | $1.7*10^7$ | $1.5*10^6$ | $1.2*10^7$ | $6.6*10^5$ | $3.1*10^6$ |

| Uncertain Level | 20% | | | | | |
|---|---|---|---|---|---|---|
| Design | 1 | 2 | 3 | 4 | 5 | 6 |
| $h_1/Z$ (spacing ratio) | 0.2884 | 0.3494 | 0.3658 | 0.3568 | 0.3398 | 0.3648 |
| $\mu$ (viscosity Pa.s) | 0.495808 | 0.171967 | 0.219008 | 0.213425 | 0.324008 | 0.234192 |
| $Z$ (stage length m) | 24.17 | 27.95 | 24.33 | 31.3 | 27.28 | 25.02 |
| $Q_o$ (injection rate $m^3/s$) | 0.23505 | 0.17493 | 0.16866 | 0.1225 | 0.17606 | 0.18011 |
| $p_{perf}$ (perforation factor $Pa.s^2/m^6$) | $1.4*10^5$ | $1.5*10^6$ | $1.3*10^5$ | $1.2*10^7$ | $1.1*10^5$ | $6.8*10^5$ |

Flow rate which shown with no obvious effect on individual design performance undertakes partial job to adjust the portfolios performance. As larger range parameters could spread, a weak dependence on the entry loss is presented in contrary to the intuition that higher perforation pressure will be required for strong heterogeneity formation. Large pressure loss could stimulate uniformly distributed flow between clusters by overwhelming the energy balance, however formation also varied in some dimensions a uniform flow rate cannot help, including elasticity, toughness and leak-off coefficient. The influence from rock parameters exceeds the energy balance, manipulating the fracture swarm. This finding will turn the trend that high perforation pressure is more and more often used in serious heterogenous reservoir. Non-necessary resource cost could be avoided though the portfolio combination at desired energy efficiency.

## 5. Conclusion

During HF, fracking fluid is injected into rock formation to create sufficient permeability, allowing significant fluid flow out as the return. Due to technique issues, the anisotropy buried in rock formation keeps unknown in large scale, driving uncertain recovery with divergent resource efficiency, which is the key concern about HF from economy and environment aspect. High-pressure injection 'extreme limited entry' is the mostly used solution but shown here with high resource cost and no capacity to overcome high formation variation. To seek other possible methods, energy efficiency is the parameter newly developed to evaluate the resource efficiency, and its' deviation between clusters is energy variance. The risk is denoted as the summation of the deviation of energy efficiency and energy variance. We recognize that the formation uncertainty along one well could be considered as an equivalent problem among multiple wells. Motivated by that, a thought to respond uncertainty by combining different designs at optimal probabilities is firstly revealed, pursuing optimization for multiple wells instead of single well. The combination is named as portfolio, of which performance estimate is based on the return energy efficiency and corresponding risk.

Massive design options ready for portfolio pick is enabled by the C5Frac, which could provide the performance of individual design in minutes with validated result. The picked design is mixed at random probability, on behalf of the proportion each design occupied in portfolio, expanding the search range to dig out the benefit of portfolio. A cloud of portfolios appear with a clear frontier, where 50% less risk is achieved at same energy efficiency with owning only one type design. The challenge is the conflict between efficiency and risk, which is hardly to achieve optimal at same time no matter the formation uncertainty. We find the portfolios formed by selected EXL designs could act the role to lower the risk of high efficiency portfolio, by mixing these two portfolios. The portfolio combinations distributed on the tangent lines drawn from the point of the low risk portfolios to the feasible region for high efficiency portfolios, where the optimal portfolio combinations located on the lowest gradient line. Compare to repeating ordinary design at multiple wells, the optimal combinations balance the risk and efficiency at best trade off with times higher energy efficiency and notably risk reduction. The advantage shown over single design relies on the reservoir condition and may be strengthen if the uncertain level could be precisely measured for each formation parameter. This work approved the innovation to get high resource efficiency with acceptable variation through probability optimized portfolio combination; revealed the possibility to weaken the environmental impact with increased recovery; indicated much room is left for further optimization on resource efficiency with more involved properties like proppants; therefore encourage the future work on formation detection especially the toughness and the uncertain level determination.

## Acknowledgments

This material is based upon work supported by the University of Pittsburgh Center for Energy, Swanson School of Engineering, Department of Chemical and Petroleum Engineering, and Department of Civil and Environmental Engineering. Additional support for recent advances to this work was provided by the National Science Foundation under Grant No. 1645246.

## References

1. Energy Information Administration. Natural Gas Gross Withdrawals and Production. 2020.
2. D. M. Kargbo, R. G. Wilhelm, D. J. Campbell, Natural Gas Plays in the Marcellus Shale: Challenges and Potential Opportunities. *Environ. Sci. Technol.* **44**, 5679-5684 (2010).
3. V. Smakhtin, C. Revenga, P. Doll, A pilot global assessment of environmental water requirements and scarcity. *Water. Int.* **29**, 307−317 (2004).
4. S. L. Brantley, R. D. Vidic, K. Brasier, D. Yoxtheimer, J. Pollak, C. Wilderman, T. Wen, Engaging over data on fracking and water quality. *Science* **359**, 395-397. (2018).
5. A. Vengosh, R. B. Jackson, N. Warner, T. H. Darrah, A. Kondash, A Critical Review of the Risks to Water Resources from Unconventional Shale Gas Development and Hydraulic Fracturing in the United States. *Environ. Sci. Technol.* **48**, 8334-8348 (2014).
6. S. Entrekin, A. Trainor, J. Saiers, L. Patterson, K. Maloney, J. Fargione, J. Kiesecker, S. Baruch-Mordo, K. Konschnik, H. Wiseman, J. P. Nicot, J. N. Ryan, Water Stress from High-Volume Hydraulic Fracturing Potentially Threatens Aquatic Biodiversity and Ecosystem Services in Arkansas, United States. *Environ. Sci. Technol.* **52**, 2349-2358 (2018).
7. W. L. Ellsworth, Injection-Induced Earthquakes. *Science* **341**, 142-149 (2013).

8. J. D. Baihly, R. Malpani, C. Edwards, SY. Han, J. C. L. Kok, E. M. Tollefsen, C.W. Wheeler, Unlocking the shale mystery: How lateral measurements and well placement impact completions and resultant production. In Proceedings SPE Tight Gas Completions Conference. San Antonio, Texas, USA. SPE 138427. 2010.

9. C. Cipolla, X. Weng, H. Onda, T. Nadaraja, U. Ganguly, R. Malpani, New algorithms and integrated workflow for tight gas and shale completions. In Proceedings SPE Annual Technology Conference and Exhibition. Denver, Colorado, USA. SPE 146872. 2011.

10. V. Sesetty, A. Ghassemi, Numerical simulation of sequential and simultaneous hydraulic fracturing. In A. P. Bunger, J. McLennan, R. G. Jeffrey, (Eds.), Effective and Sustainable Hydraulic Fracturing. Chapter 33. Rijeka, Croatia: Intech. 2013.

11. H. H. Abass, M. Y. Soliman, A. M. Tahini, J. Surjaatmadja, D. L. Meadows, L. Sierra, Oriented fracturing: A new technique to hydraulically fracture an openhole horizontal well. In Proceedings SPE Annual Technical Conference and Exhibition. New Orleans, LA, USA. SPE 124483. 2009.

12. M. K. Fisher, J. R. Heinze, C. D. Harris, B. M. Davidson, C. A. Wright, K. P. Dunn, Optimizing horizontal completion techniques in the Barnett shale using microseismic fracture mapping. In Proceedings SPE Annual Technology Conference and Exhibition. Houston, Texas, USA. SPE 90051. 2004.

13. B. Meyer, L. Bazan, A discrete fracture network model for hydraulically induced fractures- theory, parametric and case studies. In Proceedigs SPE Hydraulic Fracturing Technology Conference and Exhibition. The Woodlands, Texas, USA. SPE 140514.6. 2011.

14. C. Cheng, A. P. Bunger, A. P. Peirce. Optimal Perforation Location and Limited Entry Design for Promoting Simultaneous Growth of Multiple Hydraulic Fractures, Society of Petroleum Engineers. doi:10.2118/179158-MS. 2016.

15. J. B. Crump, M. W. Conway, Effects of perforation-entry friction on bottomhole treating analysis. *J. Petrol. Technol.* **15474**, 1041–1049 (1988).

16. A. P. Peirce, A. P. Bunger, Interference Fracturing: Non-Uniform Distributions of Perforation Clusters that Promote Simultaneous Growth of Multiple HFs. *SPE. J.* **20**, 384-395 (2015).

17. I. J. Laurenzi, G. R. Jersey, Life Cycle Greenhouse Gas Emissions and Freshwater Consumption of Marcellus Shale Gas. *Environ. Sci. Technol.* **47**, 4896-4903 (2013).

18. L. Chiapponi, V. Ciriello, S. Longo, V. D. Federico, Non-Newtonian backflow in an elastic fracture. *Water. Resour. Res.* **55**, 10144–10158 (2019).

19. A. J. Kondash, N. E. Lauer, A. Vengosh, The intensification of the water footprint of hydraulic fracturing. *Sci. Adv.* **4**, 5982-5990 (2018).

20. B. R. Scanlon, R. C. Reedy, J. P. Nicot, Will water scarcity in semiarid regions limit hydraulic fracturing of shale plays? *Environ. Res. Lett.* **9**, 124011-124025 (2014).

21. A. L. Mitchell, M. Small, E. A. Casman, Surface water withdrawals for Marcellus Shale gas development: Performance of alternative regulatory approaches in the Upper Ohio River Basin. *Environ. Sci. Technol.* **47**, 12669−12678 (2013).

22. J. H. Matthew, R. D. Heather, L. E. William, W. S. Brian, H. Chris, F. Cliff, R. O. Harrison, E. O. Jon, M. M. Beatrice, B. Casey, H. L. James, Causal factors for seismicity near Azle, Texas. *Nat. Commun.* **6**, 6728 (2015).

23. M. P. Ryan, C. C. Martin, S. J. Richard, W. Hao, High density oilfield wastewater disposal causes deeper, stronger, and more persistent earthquakes. *Nat. Commun.* **10**, 3077 (2019).

24. L. Cornelius, W. Matthew, D. Z. Mark, Physics-based forecasting of man-made earthquake hazards in Oklahoma and Kansas. *Nat. Commun*. **9**, 3946 (2018).

25. H. M. Markowitz, Portfolio Selection. *J. Finance*. **7**, 77–91 (1952).

26. C. Cheng, A. P. Bunger, Model-Based Evaluation of Methods for Maximizing Efficiency and Effectiveness of Hydraulic Fracture Stimulation of Horizontal Wells. *Geophys. Res. Lett.* **46,** 12870-12880 (2019).

27. W. Sharpe, Mutual Fund Performance. *J. Bus*. **39**, 119-138 (1966).

28. C. Cheng, A. P. Bunger, Rapid Simulation of Multiple Radially-Growing HFs Using an Energy-Based Approach. *Int. J. Numer. Int. J. Numer. Anal. Meth. Geomech*. **71**, 281–282 (2016).

29. A. A. Savitski, E. Detournay, Propagation of a penny-shaped fluid-driven fracture in an impermeable rock: asymptotic solutions. *Int. J. Solid. Struct.* **39**, 6311–6337 (2002).

30. C. Cheng, A. P. Bunger, Optimizing Fluid Viscosity for Systems of Multiple Hydraulic Fractures. *AICHE. J.* **65,** 16564 (2019).

31. C. Cheng, A. P. Bunger, Reduced order model for simultaneous growth of multiple closely-spaced radial hydraulic fractures. *J. Comput. Phys.* 376, 228-248 (2019).

32. G. R. Irwin, Analysis of stresses and strains near the end of a crack traversing a plate. *ASME. J. Appl. Mech*. **24**, 361–364 (1957).

33. M. F. Kanninen, C. H. Popelar, *Advanced Fracture Mechanics,* in *Oxford Engineering Science Series*, (Oxford Univ. Press, Oxford, UK, 1985), vol. 15.

34. G. K. Batchelor, Brownian diffusion of particles with hydrodynamic interaction. *J. Fluid. Mech*. **74**, 1–29 (1976).

35. D. Garagash, *Hydraulic fracture propagation in elastic rock with large toughness.* in Pacific Rocks 2000 'Rock around the rim', J. Girard, M. Liebman, C. Breeds, T. Doe, Eds (A.A. Balkema, Rotterdam, 2000), pp. 221-228.

36. E. Carter, in *Optimum fluid characteristics for fracture extension.* G. C. Howard, C. R. Fast, Eds. (Drilling and Production Practices, 1957), pp. 261–270.

37. I. N. Sneddon, The distribution of stress in the neighborhood of a crack in an elastic solid. *Proc. Roy. Soc.* London A **187**, 229-260 (1946).

38. A. P. Peirce, E. Detournay, An implicit level set method for modeling hydraulically driven fractures. *Comput. Method. Appl. M.* **197**, 2858-2885 (2008).

39. W. Fu, A. P. Bunger, 3D DEM Simulation on the Interference of Multiple Hydraulic Fractures in Horizontal Wells. 53rd U.S. Rock Mechanics/Geomechanics Symposium, At New York City, New York. 19–0045. 2019.

40. A. Rezaei, F. Siddiqui, G. Bornia, M. Soliman, Applications of the fast multipole fully coupled poroelastic displacement discontinuity method to hydraulic fracturing problems. *J. Comput. Phys*. **399**. 108955 (2019).

41. A. P. Bunger, Analysis of the power input needed to propagate multiple HFs. *Int. J. Solids. Struct.* **50**, 1538–1549 (2013).

42. D. Mader, Hydraulic Proppant Fracturing and Gravel Packing. Elsevier. **202**, 173–174 (1989).

43. G. Roberts, J. Whittaker, J. McDonald, T. Paxson, (2020). Proppant Distribution Observations from 20,000+ Perforation Erosion Measurements. Society of Petroleum Engineers. doi:10.2118/199693-MS